\documentclass[12pt]{article}
\usepackage[dvips]{graphicx}
\usepackage{amssymb}
\begin{document}

\newcommand{\be}{\begin{equation}}
\newcommand{\ee}{\end{equation}}

%\twocolumn[\hsize\textwidth\columnwidth\hsize\csname@twocolumnfalse\endcsname

\title{\bf Non linear flux flow in TiN superconducting thin film}

\author{F. Lefloch$^{1}$\footnote{Corresponding author, email : flefloch@cea.fr}, C. Hoffmann$^{1}$ and O. Demolliens$^{2}$
\\$^{1}$ D\'epartement de Recherche Fondamentale sur la Mati\`ere Condens\'ee
\\Service de Physique Statistique, Magn\'etisme et Supraconductivit\'e
\\$^{2}$D\'epartement des Technologies Avanc\'ees - LETI - DMEL  
\\CEA/Grenoble - 17 Avenue des Martyrs
\\38054 Grenoble cedex 9, France.}

\maketitle

\begin{abstract}

We have studied the superconducting behavior of a $100 \,nm$ Titanium Nitride (TiN) thin film in a perpendicular magnetic field. We found a zero field transition temperature of $4.6\,K$ and a slope for the transition line in the H-T plane of $-0.745\,T/K$. At $4.2\,K$, we have performed careful transport measurements by measuring both the differential resistivity and voltage as a function of a DC current. Our results are analyzed in the framework of linear and non linear flux flow behavior. In particular, we have observed an electronic instability at high vortex velocities and from its dependence with respect to the applied magnetic field, we can extract the inelastic scattering time and the diffusion length of the quasiparticles.    
\end{abstract}

\section{Introduction}

\par Vortex motion have been intensively studied over the past years, especially in high $T_{c}$ type II superconductors which show new exciting features \cite{Blatter94}. In a thin film geometry both low $T_{c}$ and high $T_{c}$ superconducting films often show similar behaviors (glass transition, vortex liquid state) and are therefore commonly used to investigate the mixed state (\cite{Hellerqvist96}-\cite{Dominguez95}). In this goal, transport properties give some insights about the motion of vortices on a macroscopic level and in some cases can be analyzed in terms of quasiparticles dynamics (\cite{Musienko80}- \cite{Xiao98}). Indeed, some 20 years ago, Larkin and Ovchinnikov (LO) predicted an electronic instability in the voltage-current characteristics at large vortex velocities \cite{LO1}, \cite{LO2}. This instability is related to the energy relaxation time of the quasiparticles $\tau_{\epsilon}$. 
\par When an electrical current passes through a sample, it creates a force on the vortices which is maximum when the direction of the current is perpendicular to the magnetic field. Above a certain value (the critical current), the vortices are depinned and start to move with a velocity proportionnal to the electric field strength E that develops along the sample. When the velocity is such that the time for a vortex to move over a distance of its size ($\approx 2\xi$) is of the order of the quasiparticles relaxation time, their number decreases inside the vortex core and increases outside. Then, the diameter of the vortex core shrinks and the viscous coefficient is reduced leading to an even higher velocity. In this regime, the I-V curves are no longer linear and at a critical velocity $v^{*}$, the differential flux flow resistivity becomes negative. For current-biased experiments an instability takes place and the system switches to another state with a measured resistivity close to the normal one. LO showed that the velocity $v^{*}$ at which the voltage jumps, is related to the relaxation time $\tau_{\epsilon}$ and should be independent of the magnetic field. 
\par However, a field dependence is often observed experimentally. There can be different reasons for this field dependence. First, in the LO theory, the diffusion length of the quasiparticles $l_{\epsilon}=\sqrt{D\tau_{\epsilon}}$ must be greater than the distance between vortices $a_{0}\sim \sqrt{\phi_{0}/B}$ because the non equilibrium distribution of the quasiparticles is assumed to be uniform over the whole superconductor volume. If this is not the case, local effects can be included by introducing explicitely the inter vortex spacing leading to a $1/\sqrt{B}$ dependence of the critical velocity $v^{*}$ \cite{Doettinger95}. In a second approach, Bezuglyj and Shklovskij (BS) refined the LO description by taking into account the temperature $T^{*}$ of the quasiparticles which can be different from that of the crystal lattice $T_{0}$ because of the finite rate of removing the power dissipated in the sample \cite{Bezuglyj92}. Again, for the BS results to be valid and used to describe experiments, the diffusion length of the quasiparticles must be larger than $a_{0}$. 
\par In this paper, we report transport measurements obtained in a $100 \,nm$ Titanium Nitride (TiN) thin film. In order to characterize the film, we first measured the temperature dependence of the resistivity for various magnetic fields applied perpendicularly to the surface of the film. This allows us to draw the phase diagramm in the H-T plane and to obtain the coherence length $\xi(0) \simeq 10 \,nm$. In a second part, we measured both the differential resistance and the voltage drop across a patterned microbridge, as a function of a DC (or slow varying) current, for different values of the magnetic field. From the differential resistance we can deduce the flux flow resistance and compare it to the LO model. From the I-V curves, we have measured the voltage instability $V^{*}$ and we show that in the field range we have investigated, the critical vortex velocity is field dependent with a $1/\sqrt{B}$ behavior. We then extracted the relaxation time $\tau_{\epsilon}$ and checked that indeed the diffusion length $l_{\epsilon}$ is comparable to the vortex distance.

\section{Sample}

\par The sample is a thin film of Titanium Nitride.It is a material that has been used for many years in microelectronics as a diffusion barrier. At low temperature, TiN can be superconductor with a critical temperature up to $\simeq 6\,K$, depending on the conditions of deposition. The film we have used is $100 \,nm$ thick and shows a  zero field critical temperature of $4.6\,K$. The film has been synthetized from a Titanium target in a mixed Argon and Nitrogen atmosphere, using a collimator. The deposition has been made on a 8 inches $Si/SiO_{2}$ wafer at a temperature of $350^{\circ}C$ with zero bias voltage. The room temperature resistivity is $85\,\mu \Omega.cm$ and is almost constant with the film thickness, from $d=100\,nm$ to $8\,nm$. Therefore, an upper bound value to the mean free path is estimated to few nanometers and seems reasonable for this kind of granular material \cite{Tsai93}. Some preleminarly results of x-rays diffraction \cite{Villegier99} show that the films are textured with two preferred orientations $<200>$ and $<111>$. Further investigations should allow us to measure the averaged grain size.    
By use of UV photolithography, we patterned two kind of bridges. Sample A was $200 \mu m$ long and $10 \mu m$ wide and sample B was $17 \mu m$ long and $7.5 \mu m$ wide. Sample A has been essentially used to measure the temperature dependence of the resistance, whereas sample B (that has a lower resistance) has been used for transport measurements at $4.2\,K$. We have checked that sample A gives similar I-V curves than sample B at 4.2 K. 

\section{Magnetic field - Temperature diagram}

\begin{figure}
\begin{center}
\includegraphics[width=10cm]{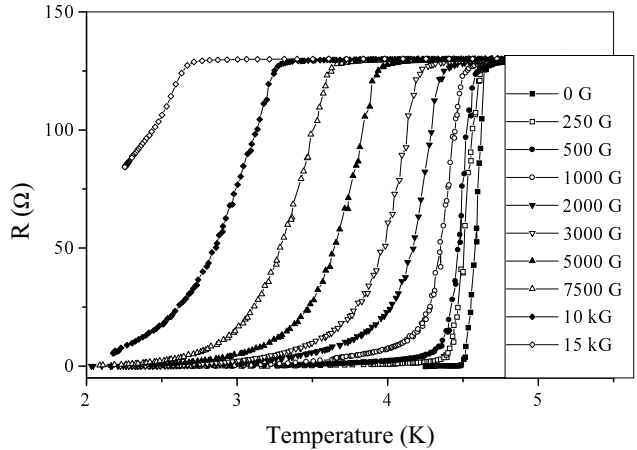}
\caption{\footnotesize Temperature dependence of the resistance of Sample A for various perpendicular magnetic fields.}
\label{fig:r_t}
\end{center}
\end{figure} 

\begin{figure}
\begin{center}
\includegraphics[width=10cm]{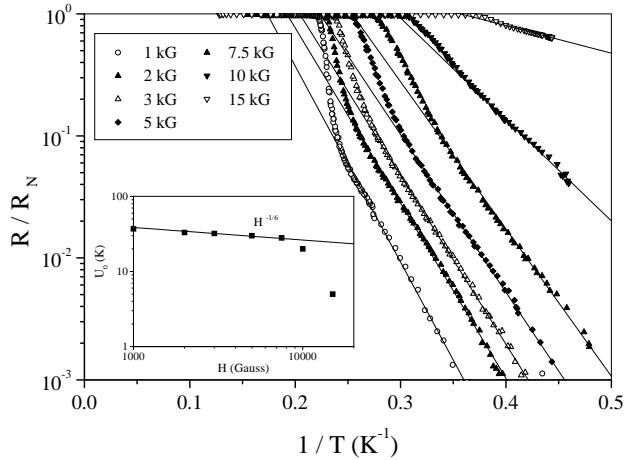}
\caption{\footnotesize Log-lin plot of the relative resistance drop around $T_{c}$ as a function of the inverse temperature for several magnetic fields from $1\,kG$ to $15\,kG$. In insert, the field dependence of the energy scale of the thermally activated flux flow regime.}
\label{fig:tintaff}
\end{center}
\end{figure}

\par
\begin{figure}
\begin{center}
\includegraphics[width=10cm]{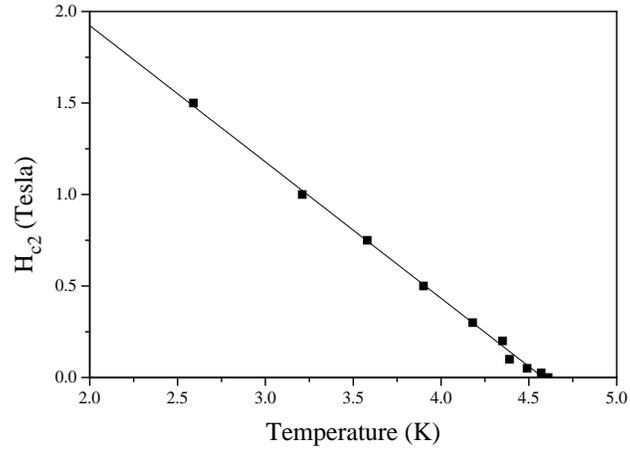}
\caption{\footnotesize Critical magnetic field versus critical temperature. The slope is $-0.745\, T/K$ which gives $\xi(0)\simeq 10\,nm$ (See text).}
\label{fig:h_t}
\end{center}
\end{figure}

 Figure \ref{fig:r_t} shows the resistance of Sample A as a function of temperature for different magnetic fields. In zero field, the transition width is about $0.1\,K$.  We see that the transition becomes larger as the magnetic field increases, especially at the foot of the transition. This broadening is due to thermally activated flux flow as it can be seen in figure \ref{fig:tintaff}. Indeed, over a certain range of temperature, the resistance decrease can be described by a thermally activated behavior with a field dependent energy scale (see insert of figure \ref{fig:tintaff}). We did not investigate this behavior any further but is similar to what has been observed by Hsu et al. \cite{Hsu92}. Since, the transition temperature is not very well defined at high field, we took as a criterion for $T_{c}$, the temperature at which the resistance drops to $90\,\%$ of its value in the normal state. This will overevaluate the transition temperature but will not change the slope in the H-T plane. Indeed, contrary to the foot, the shape of the beginning of the transition does not change much with the field.
Figure \ref{fig:h_t} shows the critical magnetic field $H_{c2}$ versus the critical temperature $T_{c}$. The slope ${\left (\partial H_{c2} \over{\partial T} \right )}_{H=0}$ is $-0.745 \,T/K$ and gives $\xi (0) \simeq \,10\,nm$ using $\xi (T)=\xi (0)/\sqrt{1-T/T_{c}}$ and $H_{c2}=\Phi_{0}/2\pi\xi^{2}$

\section{Transport measurements}

\begin{figure}
\begin{center}
\includegraphics[width=10cm]{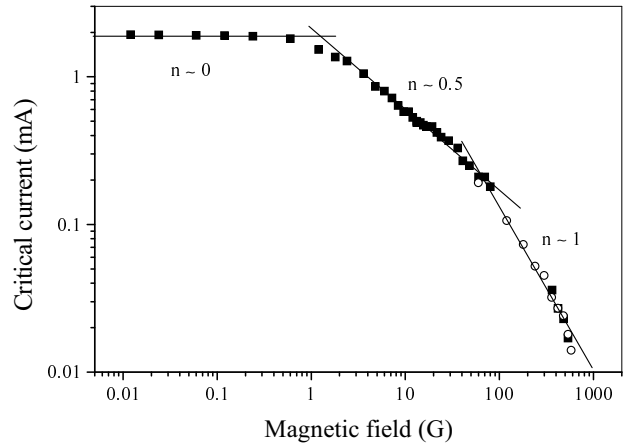}
\caption{\footnotesize Log-log plot of the critical current versus magnetic field. The critical current has been obtained from I-V curves (solid symbol) and from differential resistance measurements versus DC current (open symbol).}
\label{fig:ic_h}
\end{center}
\end{figure}

\par For these experiments, the magnetic field was generated by a 650 turns NbTi superconducting coil. Both sample and coil were immersed in the Helium bath of a magnetically unshielded dewar. The currents (DC and AC) were generated by a voltage driving current source following the electronic scheme of Payet-Burin et al. \cite{PPB}. The voltage was amplified by a low noise differential preamplifier. 

\subsection{Critical current}

We have measured the onset of dissipation by measuring I-V curves at very low voltage. In Figure \ref{fig:ic_h}, we have plotted the current at which the voltage exceeds $\simeq 1\,\mu V$ across the sample B and identified it to the critical current. On a log-log plot, we see that below a certain field of few Gauss, the critical current $I_{c}=1.9 \,mA$ and is constant in field over two decades. For such a current flowing in the film, the self-magnetic field generated by the current at the edge of the bridge is : $B_{sf}={\mu_{0}I\over 2 \pi w }ln{\left ({2w\over d} \right)} \simeq 2\,G$ for a width $w=10\mu \,m$ and a thickness $d=100\,nm$. Therefore for field less than $\simeq 2 \,G$ the main "applied" field is the self-field. For higher fields, the observed behavior is consistent with a regime of collective pinning and the critical current decreases as a power law of the field $I_{c} \propto B^{-n}$. Two regimes can be distinguished with $n\simeq 0.5$ and $n\simeq 1$ but their origins are not clear. This behavior can be related to that observed by De Brion et al. \cite{Debrion94} in $Mo_{77}Ge_{23}$.  

\subsection{Core resistivity}

\begin{figure}
\begin{center}
\includegraphics[width=10cm]{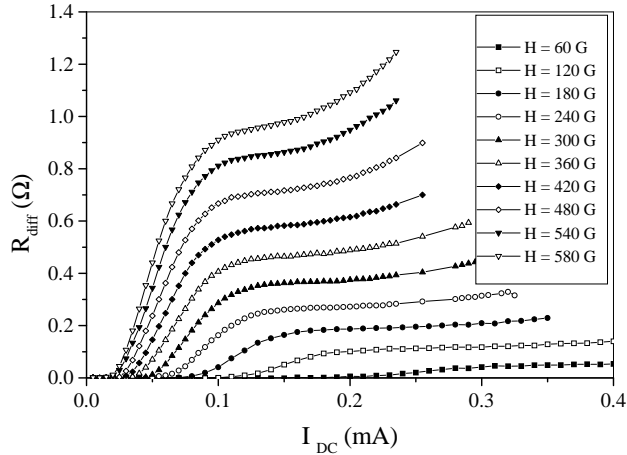}
\caption{\footnotesize Differential resistance versus DC current in sample B immersed in liquid helium at $4.2\,K$.}
\label{fig:rdiff}
\end{center}
\end{figure}

When measuring the transport behavior in the dissipative regime ($I>I_{c}$), one can distinguish different behaviors. Those appear more clearly by recording the differential resistance $R_{diff}={\partial V\over \partial I}$ versus a DC current. For these measurements, the amplitude of the small AC current added to the DC one, was $4\,\mu A$ peak to peak and its frequency around 2 kHz (see figure \ref{fig:rdiff}). Just above the critical current the differential resistance increases in a short DC current range. Then $R_{diff}$ crosses over a plateau before growing up again. We attribute these different regimes to the flux creep, flux flow and non-linear flux flow, respectively. We did not study the flux creep regime but we checked the flux flow regime by recording the value of the differential resistance at the plateau as a function of the applied magnetic field. According to LO and \cite{Samoilov95}, the flux flow resistivity is given by : 
\be
\rho_{ff}={\left (\partial V\over \partial I \right )}_{V=0}={\tilde{\rho}_{N}\over \alpha (B)+1}
\label{eq:rhon}
\ee
with
\be
\alpha (B) = {1 \over \sqrt{1-{T\over T_{c}}}}{B_{c2}\over B}f{\left (B\over B_{c2} \right)}  
\ee
and 
\be
f{\left (B\over B_{c2} \right )}=4.04-{\left (B\over B_{c2}\right )}^{1/4}{\left ( 3.96+2.38{B\over B_{c2}} \right )} \quad \textrm{when} \quad {B\over B_{c2}} \lesssim 0.315 
\ee     
where $\tilde{\rho}_{N}$ is the normal core resistivity and $B_{c2}$ the critical field at 4.2 K ($B_{c2} = 3 \,kG$ from figure \ref{fig:h_t}). The above expressions are valid for temperatures close to $T_{c}$ and are therefore well suited to describe our experiments since a temperature of $4.2\,K$ corresponds to $0.9\,T_{c}$.
\par
\begin{figure}
\begin{center}
\includegraphics[width=10cm]{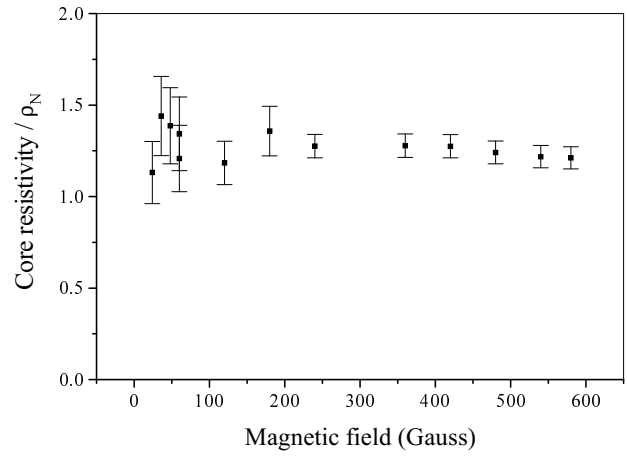}
\caption{\footnotesize Normal core resistivity / normal sample resistivity ratio versus applied magnetic field, at $4.2\,K$.}
\label{fig:rn}
\end{center}
\end{figure}

 In figure \ref{fig:rn}, we have plotted the normal core resistivity deduced from the measurements of the resistivity at the plateau and following equation \ref{eq:rhon}. One can see that the core resistivity is indeed independent of the field, but shows a value slightly above the normal resistivity of the sample at $T > T_{c}$. This kind of discrepancy has been already reported by Doettinger et al. \cite{Doettinger96} and can be due to some uncertainties on the coefficients in the above formula or to some differences between the resistivity of the vortex core and the normal resistivity of the whole sample.
\par In the flux flow regime, the differential resistivity is constant as a function of the DC current. Looking at our experimental results, this is clearly not the case (see figure \ref{fig:rdiff}) and the system enters the non-linear flux flow regime.

\subsection{Non linear flux flow and electronic instability}

\begin{figure}
\begin{center}
\includegraphics[width=10cm]{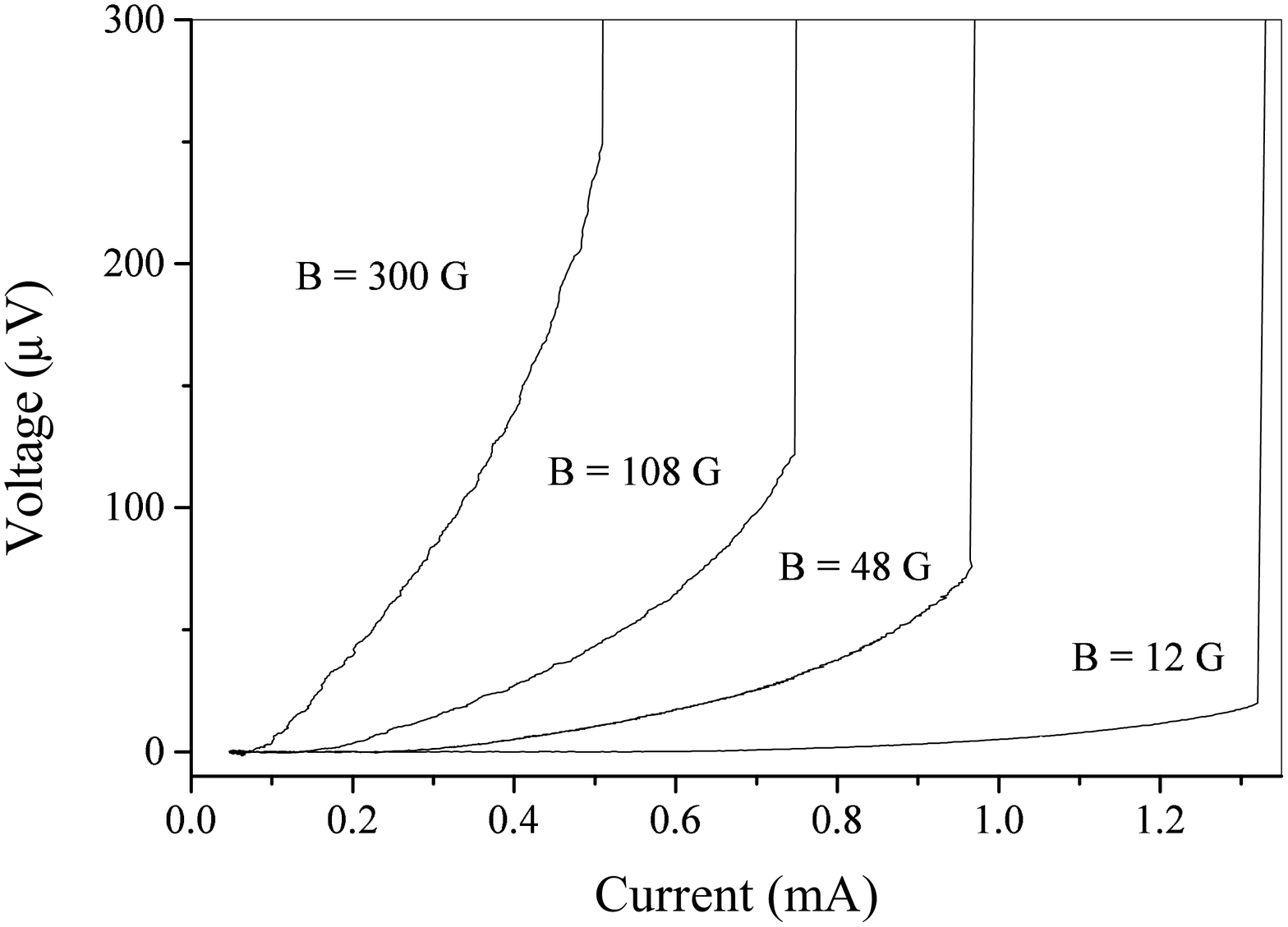}
\caption{\footnotesize Voltage-current characteristics at $4.2\,K$ under various magnetic fields.}
\label{fig:v_i2}
\end{center}
\end{figure}
\par
\begin{figure}
\begin{center}
\includegraphics[width=10cm]{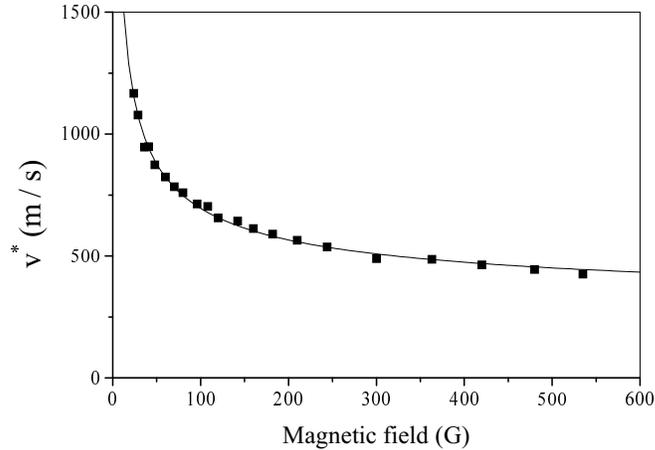}
\caption{\footnotesize Plot of the critical vortex velocity obtained from the IV curves versus magnetic field. The solid line follows the $1/\sqrt{B}$ behavior expected when the distance between vortices is not small relatively to the quasiparticles diffusion length $l_{\epsilon}$. }
\label{fig:vstar}
\end{center}
\end{figure}

 In order to analyse the regime of non linear I-V curves, we have studied the voltage response versus a slow varying current (figure \ref{fig:v_i2}). To avoid overheating of the sample by Joule effect, we have used triangular current pulses of few ms with a duty cycle of about 10 (i.e the pulse was repeated every 100 ms with zero current in between). The voltage response was recorded on a digital oscilloscope and averaged more than 30 times. Figure \ref{fig:v_i2} shows several IV characteristics that have been recorded this way. At a voltage $V^{*}$ and current $I^{*}$ and instability takes place. We have checked that the rapid jump of the voltage at $V^{*}$ is not due to some overheating of the sample by recording the power $P=V^{*}I^{*}$ as a function of the field. We found that $P$ increases smoothly from $50\, nW$ at $20\,G$ to $150 \,nW$ at $600\,G$. As mentionned by Xiao et al. \cite{Xiao99} this power would have been constant if the main mechanism for the observed instability was due to heating.     
From LO model, the voltage-current characteristics behave as :
\be
I-I_{c}\simeq I \simeq {V\over \tilde{R}_{N}}{\left ({\alpha (B)\over 1+{\left ({V\over V^{*}}^2 \right )}}+1 \right )} 
\label{eq:iv}
\ee
where $V^{*}$ ($I^{*}$) is the voltage (current) at which the electronic instability takes place. Equation \ref{eq:iv} gives the correct asymptotic, and observed experimentally, behaviors : As $V\to 0$ the characteristic is linear with a constant flux flow resistance and for $V\gg V^{*}$, the I-V response is again linear with a resistance close to the normal resistance. According to LO, the voltage $V^{*}$ should be such that the vortex velocity $v^{*}=V^{*}/BL$ is constant with respect to the applied magnetic field. If one plots the critical velocity, its behavior as a function of the field is not constant as shown in figure \ref{fig:vstar} but is a decreasing function of the magnetic field. According to Doettinger et al. \cite{Doettinger95}, the vortex velocity increases at low field, in order to keep the distance $v^{*}\tau_{\epsilon}$ large enough to ensure spatial homogeneity of the quasiparticles distribution. This condition, which is essential for the use of the LO description, is achieved when $v^{*}\tau_{\epsilon}$ is comparable to the vortex spacing $a_{0}$. Therefore, the velocity $v^{*}$ is related to the magnetic field by : 

\be
v^{*}(B)=a_{0}{f(T)\over \tau_{\epsilon}}=\sqrt{{2\over \sqrt{3}}{\phi_{0}\over B}} {f(T)\over \tau_{\epsilon}}
\label{eq:v_b}
\ee  
with
\be
f(T)\simeq 1.14{\left (1-{T\over T_{c}} \right )}^{1/4}
\label{eq:f(T)}
\ee

At large field, $v^{*}$ reaches the field independent LO value :

\be
v^{*}_{LO}=\sqrt{{D\over \tau_{\epsilon}}}1.14{\left (1-{T\over T_{c}} \right )}^{1/4}
\label{eq:vlo}
\ee

where D is the quasiparticles diffusion constant ($D=1/3\nu_{F}l$, with $\nu_{F}$ the Fermi velocity and $l$ the electron mean free path). In figure \ref{fig:vstar}, the solid line is a fit using :

\be
v^{*}=v^{*}_{LO}{\left (1+{a_{0}\over\sqrt{D\tau_{\epsilon}}} \right )}
\label{vtot}
\ee
The diffusion constant $D$ is known from the slope ${\partial H_{c2}\over \partial T_{c}}$ by :

\be
D={4k_{B}\over \pi e}{\partial H_{c2}\over \partial T_{c}}^{-1}=1.26\,10^{-4}\,m^{2}/s
\ee

From this analysis, we find : $\tau_{\epsilon}\simeq 5\,10^{-10}s$. This value is in the same magnitude range than those obtained in other thin film materials at $0.9\,T_{c}$ : $\tau_{\epsilon}\simeq 10^{-11}\,s$ in $YBa_{2}Cu_{3}O_{7-\delta}$ \cite{Doettinger94}, $\tau_{\epsilon}\simeq 10^{-10}\,s$ in $Bi_{2}Sr_{2}CaCu_{2}O_{8-\delta}$ \cite{Xiao98} and $\tau_{\epsilon}\simeq 10^{-9}\,s$ in $Mo_{3}Si$ \cite{Doettinger97}. From our results, we can get the quasiparticles diffusion length $l_{\epsilon}=\sqrt{D\tau_{\epsilon}}=250\,nm$. We then check that indeed the quasiparticles diffuse over a distance which is not large compared to the inter vortex spacing $a_{0}$ which ranges from $1.5\mu\,m$ at $10\, G$ to $0.2 \mu \,m$ at $600\, G$

\section{Conclusion}
In this paper, we have given some properties of the superconducting phase of a Titanium Nitride (TiN) thin film ($100 \,nm$). We have drawn the transition line in the H-T plane with the magnetic field perpendicular to the film. From the slope ${\partial H_{c2}\over \partial T_{c}}$ between $T_{c}(0)=4.6\,K$ and $2.5\,K$, we estimate the coherence length $\xi (0)\simeq 10\,nm$. At $4.2\,K$, we performed transport measurements by recording both the differential resistance versus DC current and I-V characteristics for various field amplitudes. At low current, flux flow theory applies whereas at higher current the behavior is non-linear. For a certain current, an electronic instability takes place which corresponds to a critical vortex velocity. From its behavior with respect to the field, we get an estimate of the energy relaxation time : $\tau_{\epsilon}\simeq 5\,10^{-10}s$. Further measurements, especially at different temperatures and film thicknesses, are needed to give a more precise description of the microscopic dynamics in this material, in particular concerning electron-electron and electron-phonon relaxation processes. 

\section{Acknowledgment}
We would like to thank J. C. Vill\'egier and R. Calemczuk for clarifying discussions.

\end{document}